\documentclass[numbers]{elsarticle}
\usepackage{latexsym}
\usepackage{natbib}
\usepackage[latin2]{inputenc}
\usepackage[hidelinks]{hyperref}
\usepackage{longtable}

\makeatletter
\def\ps@pprintTitle{%
	\let\@oddhead\@empty
	\let\@evenhead\@empty
	\def\@oddfoot{\centerline{\thepage}}%
	\let\@evenfoot\@oddfoot}
\makeatother

\newtheorem{theorem}{Theorem}

\newtheorem{definition}[theorem]{Definition}

\begin{document}

\title{Comparative Connectomics: Mapping the Inter-Individual Variability of Connections within the Regions of the Human Brain}

\author[p]{Csaba Kerepesi}
\ead{kerepesi@pitgroup.org}
\author[p]{Balázs Szalkai}
\ead{szalkai@pitgroup.org}
\author[p]{Bálint Varga}
\ead{balorkany@pitgroup.org}
\author[p,u]{Vince Grolmusz\corref{cor1}}
\ead{grolmusz@pitgroup.org}
\cortext[cor1]{Corresponding author}
\address[p]{PIT Bioinformatics Group, Eötvös University, H-1117 Budapest, Hungary}
\address[u]{Uratim Ltd., H-1118 Budapest, Hungary}

\date{}

\begin{abstract}
The human braingraph, or connectome is a description of the connections of the brain: the nodes of the graph correspond to small areas of the gray matter, and two nodes are connected by an edge if a diffusion MRI-based workflow finds fibers between those brain areas. We have constructed 1015-vertex graphs from the diffusion MRI brain images of 392 human subjects and compared the individual graphs with respect to several different areas of the brain. The inter-individual variability of the graphs within different brain regions was discovered and described. We have found that the frontal and the limbic lobes are more conservative, while the edges in the temporal and occipital lobes are more diverse. Interestingly, a ``hybrid'' conservative and diverse distribution was found in the paracentral lobule and the fusiform gyrus. Smaller cortical areas were also evaluated: precentral gyri were found to be more conservative, and the postcentral and the superior temporal gyri to be very diverse.
\end{abstract}

\maketitle

\section{Introduction} 

 Large co-operative research projects, such as the Human Connectome Project \cite{McNab2013}, produce high-quality MRI-imaging data of hundreds of healthy individuals. The comparison of the connections of the brains of the subjects is a challenging problem that may open numerous research directions.  In the present work we map the variability of the connections within different brain areas in 392 human subjects, in order to discover brain areas with higher variability in their connections or other brain regions with more conservative connections. 
 
 The braingraphs or connectomes are the well-structured discretizations of the diffusion MRI imaging data that yield new possibilities for the comparison of the connections between distinct brain areas in different subjects \cite{Ingalhalikar2014b, Szalkai2015} or for finding common connections in distinct cerebra \cite{Szalkai2015a}, forming a common, consensus human braingraph.  
 
 Here, by using the data of the Human Connectome Project \cite{McNab2013}, we describe, by their distribution functions, the inter-individual diversity of the braingraph connections in separate brain areas in 392 healthy subjects of ages between 22 and 35 years. 
 
 Since every brain is unique, the workflow that produces the braingraphs consists of several steps, including a diffeomorphism \cite{Hirsch1997} of the brain atlas to the brain-image processed. After the diffeomorphism, corresponding areas of different human brains are pairwise identified through the atlas and, consequently, can be compared with one another. The braingraphs, with nodes in the corresponded brain areas, are prepared from the diffusion MRI images of the individual cerebra through a workflow detailed in the ``Methods'' section. Every braingraph studied contains 1015 nodes (or vertices). The vertices correspond to the subdivision of anatomical gray matter areas in cortical and subcortical regions. For the list of the regions and the number of nodes in each region, we refer to Table S1 and Figure S1 in the Appendix. 
 
 Next, we describe the variability, or the distribution of the graph edges in each brain region, and also in each lobe. Figure 1 contains a simplified example on three small graphs (1,2,3) each with only two regions (A \& B). The example clarifies the method, the way the results are presented through a distribution function, and the diagrams describing these functions. 
 
 For any fixed brain area, and for any $x: 0\leq x\leq 1$, let $F(x)$ denote the fraction of the edges\footnote{i.e., the number of the edges in question, divided by the number of all edges in the fixed area;} in the fixed area\footnote{i.e., with both vertices in the fixed area;} that are present in at most the fraction $x$ of all braingraphs, (for a more exact definition of $F(x)$ we refer to the ``Methods'' section). We note that $F(x)$ is a cumulative distribution function \cite{Feller2008} of a random variable described in the ``Methods'' section.

 \begin{figure}[h!]
 	\centering
 	\includegraphics[width=140mm]{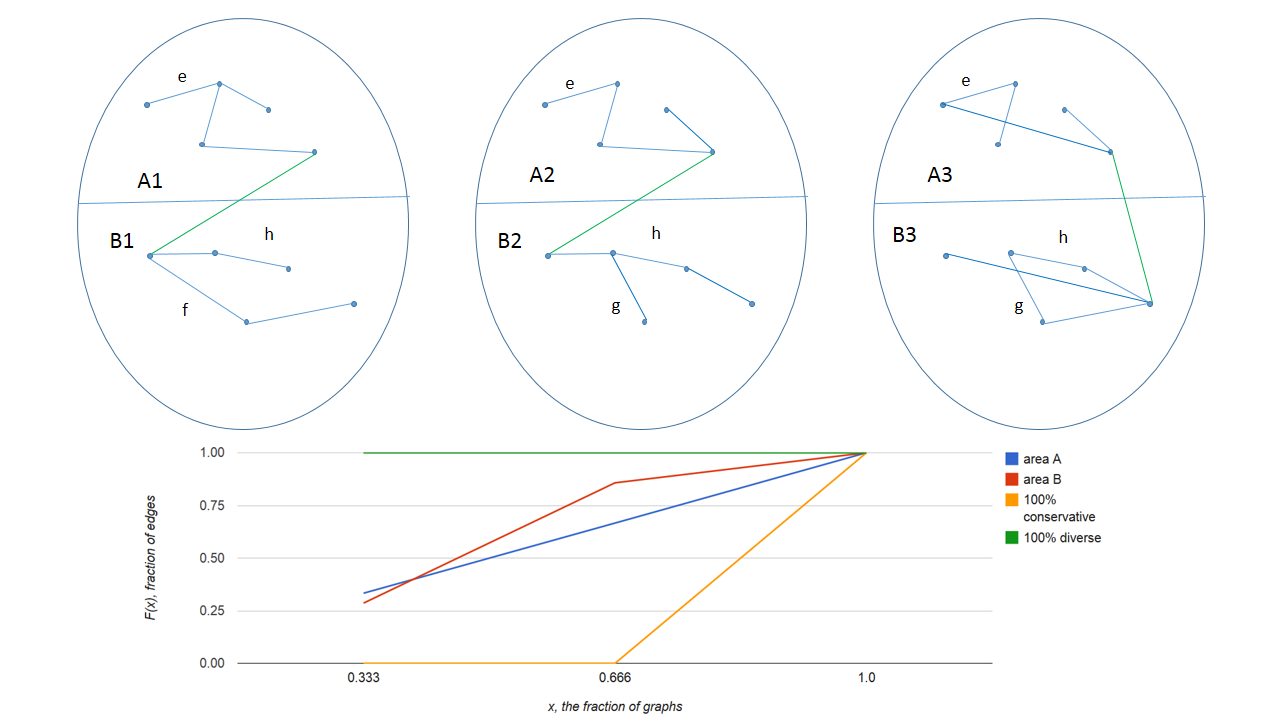}
 	\caption{A simple example of computing the edge distribution between brain areas. In the example, there are three ``braingraphs'', each with two areas: $A$ and $B$. We intend to count the edges that are present in all three graphs, only in two graphs and only in a single graph, respectively (between the same nodes, but in different graphs). For example, the copies of edge $e$ are present in all three $A$ areas, copies of edge $h$ in all three $B$ areas, copies of edge $g$ in two $B$ areas and edge $f$ is present only in $B1$. The edges crossing the boundary of $A$ and $B$ (colored green) are ignored when counting the edge distribution within the areas $A$ and $B$. In area $A$, two edges are present once, two edges twice and also two edges (including edge $e$) exactly three times. In area $B$, two edges (including $f$) are present once, four edges (including $g$) twice and one edge -- $h$ -- three times. In the diagram on the bottom, we give the $F(x)$ distribution functions for both areas. On axis $x$, the fractions of the graphs are given, 1/3 correspond to one graph, 2/3 for two and 1.0 for all three graphs. $F(x)$ is defined as the fraction of the edges in the fixed area that are present in at most the fraction $x$ of all braingraphs. Data points corresponding to area $A$ are on the same blue line (1/3, 2/3, 1) and those, corresponding to area $B$ are on the broken, red line (2/7, 6/7, 1). We remark that if all three graphs are the same, then the data points are (0,0,1) (the extremely conservative case, orange line). Similarly, if no two graphs have the same edges, the data points are (1,1,1) (that is the extremely diverse case, green line).  This type of diagram is used for the presentation of the results of the distribution of the edges in separate areas of the brain: The faster the line reaches the top $F(x)=1$ value, the more diverse is the edge set in the corresponding brain area. We also note that in the diagram the lines connect the data points corresponding to the discrete values on axis $x$, and {\em do not} describe the step-function $F(x)$ {\em between} the data points: we have chosen this visualization method because of its clarity even if a higher number of areas are shown (c.f. Figures 2 and 3 with numerous crossing lines).}\label{egy} 
 \end{figure}

\section{Results and Discussion}

Table 1 summarizes the edge diversity results for the 392 graphs for the lobes of the brain, described by the distribution functions $F(x)$. The last column contains the data for the whole brain with 1015 nodes and 70,652 edges. The sum of the edges of the lobes in Table 1 is 30,326: these edges have both endpoints in the same lobe. More than forty thousand edges are present and accounted for only in the last column, because these edges connect nodes from different lobes. Therefore, the values in the last column cannot be derived from the other columns, since that column contains the contribution of edges that do not contribute to any other columns.

We want to find out which brain areas are more conservative and which are more diverse than the others. 
We suggest to designate an area as ``conservative'' if for most $x$ values, its $F(x)$ distribution function is less than the $F(x)$ of the all brain, given in the last column. We also suggest to designate an area as ``diverse'' if for most $x$ values, its $F(x)$ distribution function is greater than the $F(x)$ of the all brain, given in the last column.

The most conservative lobes are the smallest ones: the brainstem, the thalamus and the basal ganglia contain only 1, 2 and 8 nodes, resp., and most of the edges in those regions are present in almost all braingraphs. If we take the average number of the braingraphs containing an edge from those regions, we get 316, 390 and 213 graphs, resp. 

It is much more interesting to review the diversity of the connections in larger areas. The frontal and the limbic lobes are conservative for most values of $x$ (i.e., their $F(x)$ values are less than that of the last column), while the temporal and the occipital lobes are diverse for larger $x$'s. The distribution of the edges in the fusiform gyrus is particularly interesting: more than 10\% of the graphs contain 46\% of the edges which means this is a conservative brain area in that parameter domain, compared to the other lobes. The fusiform gyrus remains conservative for $x=0.2$ and even for $x=0.3$, but more than 50\% of the graphs contain only 0.7\% of the edges. That means that some edges of the fusiform gyrus are well conserved, and some parts are very diverse. The paracentral lobule has a very similar distribution.

\begin{table}[h!]
	\centering
	\includegraphics[width=130mm]{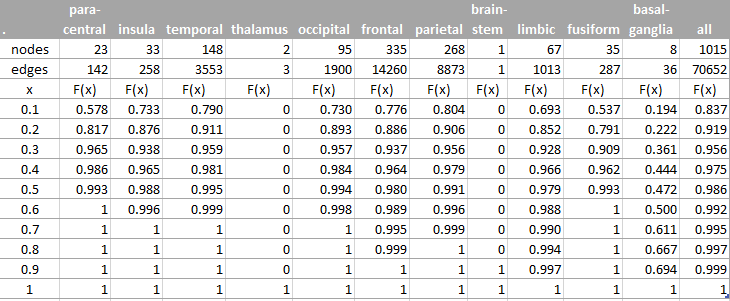}
	\caption{The number of nodes, the number of edges and the diversity of the edges in different lobes, measured by the distribution function $F(x)$. The list includes some brain areas that usually are not counted as lobes: like the fusiform gyrus, basal ganglia, and the paracentral lobule. The lobes, whose columns reach the value 1 faster (i.e. have more 1's at the bottom) have higher diversity. For example, the frontal and the limbic lobes are more conservative, while the temporal and the occipital lobes are more diverse. The distribution of the edges in the fusiform gyrus is particularly interesting: more than 10\% of the graphs contain 46\% of the edges which means this is a conservative brain area in that parameter domain, compared to the other lobes. The fusiform gyrus remains conservative for $x=0.2$ and even for $x=0.3$, but more than 50\% of the graphs contain only 0.7\% of the edges. Therefore, some edges of the fusiform gyrus are well conserved, and some other parts are very diverse. The paracentral lobule has a similar distribution. The data are also visualized on Figure 2 and an interactive figure \url{http://uratim.com/diversity/Figure_2.html}  }
\end{table}

\begin{figure}[h!]
	\centering
	\includegraphics[width=130mm]{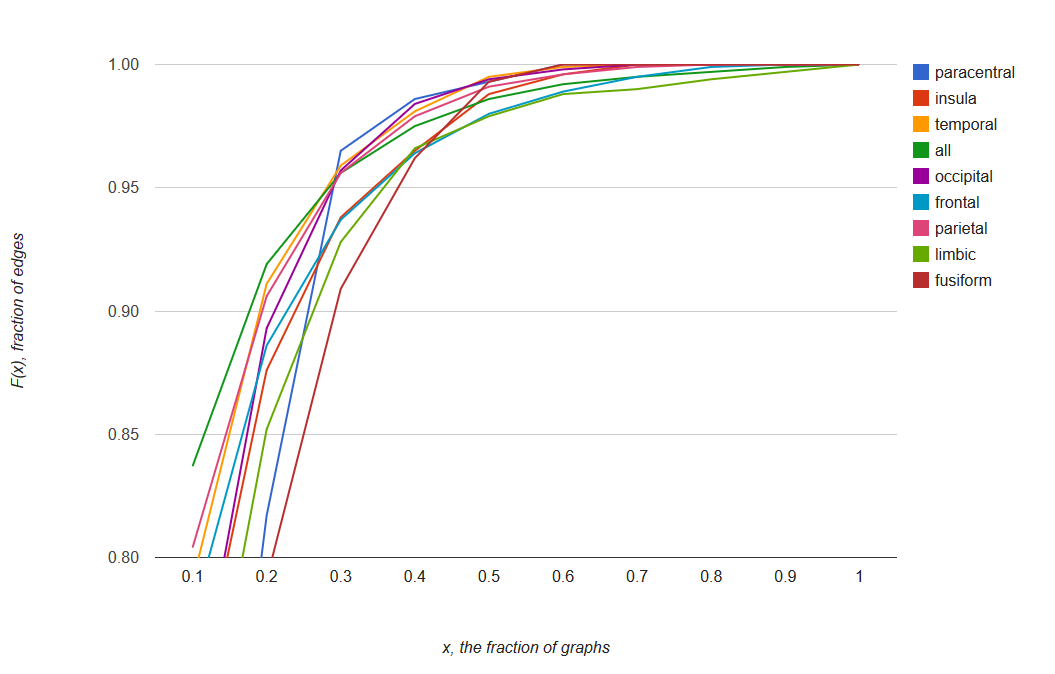}
	\caption{The diversity of the edges in different lobes, measured by the distribution function $F(x)$. Only the areas with more than 10 nodes and $F(x)$ values of more than 0.8 are visualized. The lobes, whose lines faster (i.e., with smaller $x$) reach value 1, have higher diversity. 			
		 The fusiform gyrus and the paracentral lobule clearly moves from the bottom to the top of the diagram, relative to the other lines: this observation suggests that some of their edges are very conservative, and other areas have high diversity.  An interactive version of this figure can be found at \url{http://uratim.com/diversity/Figure_2.html}  }
\end{figure}

Table 2 summarizes the diversity results for those cortical areas which have more than 222 edges (see Table S2 in the Appendix for the edge numbers).

\begin{table}[h!]
	\centering
	\includegraphics[width=130mm]{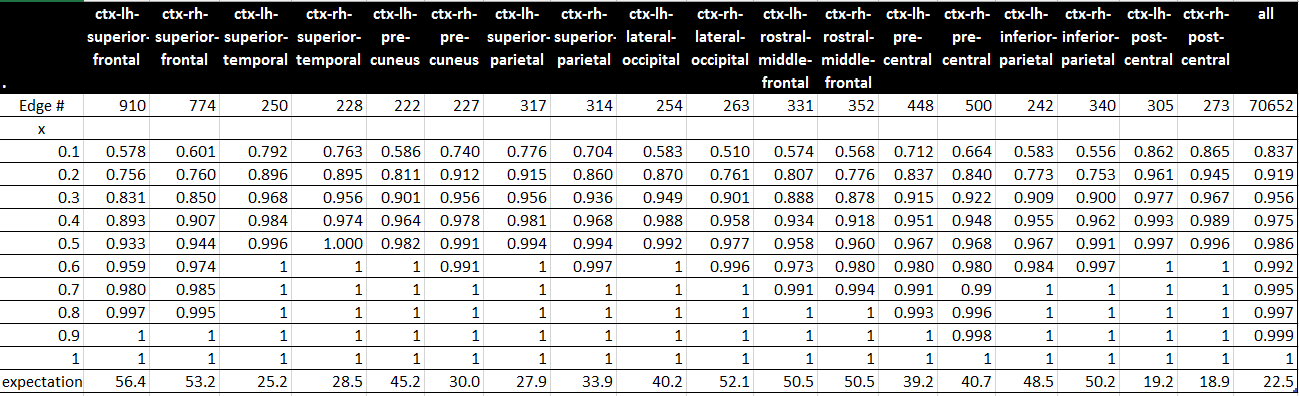}
	\caption{The diversity of the edges in different cortical areas, measured by the distribution function $F(x)$. The abbreviation ``ctx-lh'' stands for ``cortex left-hemisphere'', ``ctx-rh'' for ``cortex right-hemisphere''. The areas, whose columns reach the value 1 faster (i.e., have more 1's at the bottom)  have higher diversity. As in Table 1, the frontal regions are relatively more conservative, while the parietal regions are more diverse. Both precentral gyri are also conservative, and the postcentral and the superiortemporal gyri are more diverse. 
	The last row contains the expected number of the graphs which contain a randomly chosen edge from the brain area indicated. Large expected number implies a conservative area, a small value implies a more diverse area. The data for the left hemisphere are also visualized on Figure 3 and on an interactive figure \url{http://uratim.com/diversity/Figure_3.html}  }
\end{table}

\begin{figure}[h!]
	\centering
	\includegraphics[width=130mm]{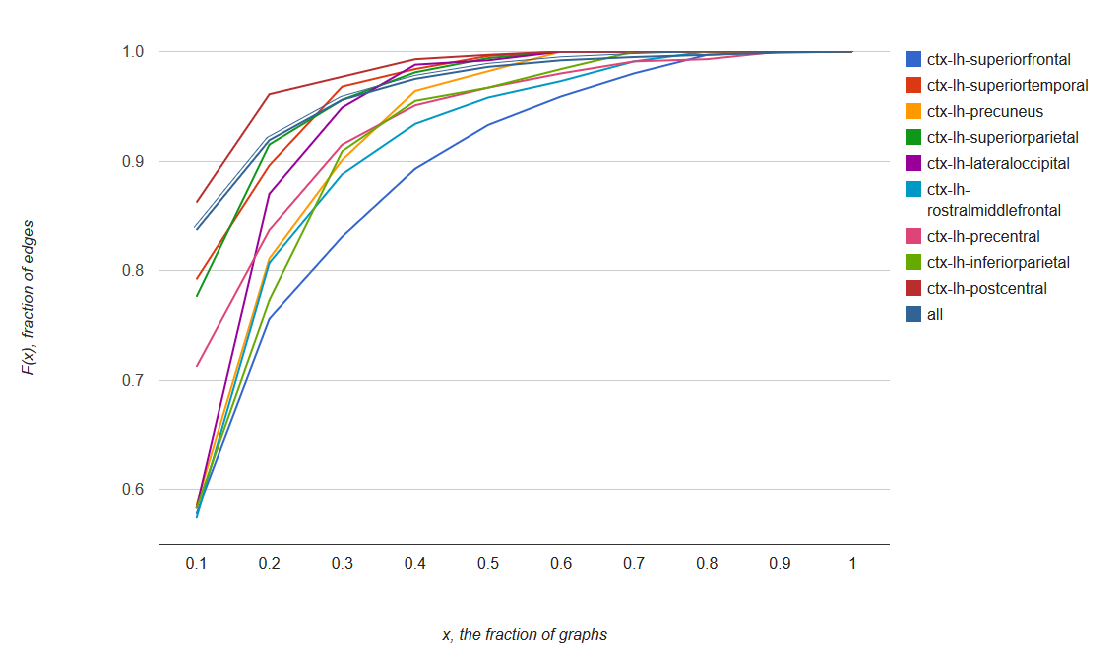}
	\caption{The diversity of the edges in different cortical areas of the left hemisphere, measured by the distribution function $F(x)$. The areas, whose lines faster (i.e., with smaller $x$) reach value 1, have higher diversity. 			
		 An interactive version of this figure can be found at \url{http://uratim.com/diversity/Figure_3.html}  }
\end{figure}

\section{Methods} 

We have worked with a subset of the anonymized 500 Subjects Release published by the Human Connectome Project \cite{McNab2013}:
\noindent (\url{http://www.humanconnectome.org/documentation/S500}) of healthy subjects between 22 and 35 years of age. Data were downloaded in October, 2014.

We have applied the Connectome Mapper Toolkit \cite{Daducci2012} (\url{http://cmtk.org}) for brain tissue segmentation, partitioning, tractography and the construction of the graphs. The fibers were identified in the tractography step. The program FreeSurfer was used to partition the images into 1015 cortical and sub-cortical structures (Regions of Interest, abbreviated: ROIs), and was based on the Desikan-Killiany anatomical atlas \cite{Daducci2012}(see Figure 4 in \cite{Daducci2012}). Tractography was performed by the Connectome Mapper Toolkit \cite{Daducci2012}, using the MRtrix processing tool \cite{Tournier2012} and choosing the deterministic streamline method with randomized seeding.

The graphs were constructed as follows: the 1015 nodes correspond to the 1015 ROIs, and two nodes were connected by an edge if there exists at least one fiber connecting the ROIs corresponding to the nodes.

\subsection{The distribution function} The variability of the edges in regions or lobes are described by cumulative distribution functions (CDF) (also called just the ``distribution function'') of the edges \cite{Feller2008}.  The general definition of the CDF is as follows:

\begin{definition}
	Let $Y$ be a real-valued random variable. Then 
	$$F(x)=Pr(Y\leq x) $$
	defines the cumulative distribution function of $Y$ for real $x$ values.
\end{definition}

For example, if $a$ is the maximum value of $Y$ then $F(a)=1$, and if $b$ is less than the minimum value of $Y$, then $F(b)=0$.

CDFs are used the following way: Suppose that our cohort consists of $n$ persons' braingraphs (in the present work $n=392$). For a given, fixed brain area, our random variable $Y$ takes on values $Y=u/n, u=0,1,\ldots,n$. The equation $Y=u/n$ corresponds to the event that a uniformly, randomly chosen edge is in exactly $u$ graphs from the $n$ possible one, and the probability $Pr(Y=u/n)$ gives the probability of this event. Or, in other words, the equation $Y=u/n$ corresponds to the set of edges --- with both nodes in the fixed brain area --- which are present in exactly $u$ braingraphs, and the probability $Pr(Y=u/n)$ gives the fraction of the edges that are present in exactly $u$ braingraphs. Therefore, $F(x)=Pr(Y\leq x)$ gives the fraction (i.e., the probability) of the edges that are present in at most of a fraction $x$ of all the graphs.

The number of nodes and edges in each brain regions are given in supporting Tables S1 and S2 in the Appendix. We remark that we counted the edges without multiplicities: that is, if an edge $e$ was either present in, say, 42 copies or just 1 copy of the braingraph, in both cases we counted it only once.

The distributions were computed by counting the number of appearances of each edge in all the 392 braingraphs. Then the distribution of these numbers were evaluated in lobes and smaller cortical areas.

\section{Conclusions:} By our knowledge for the first time, we have mapped the inter-individual variability of the braingraph edges in different cortical areas. We have found more and less conservative areas of the brain: for example, frontal lobes are conservative, superiortemporal and  the post-central gyri are very diverse. The fusiform gyrus and the paracentral lobule have shown both conservative and diverse distributions, depending on the range of the parameters. 

\section*{Data availability:} The unprocessed and pre-processed MRI data are available at the Human Connectome Project's website:

http://www.humanconnectome.org/documentation/S500 \cite{McNab2013}. 

\noindent The assembled graphs that were analyzed in the present work can be accessed and downloaded at the site 

\url{http://braingraph.org/download-pit-group-connectomes/}.

\section*{Acknowledgments}
Data were provided in part by the Human Connectome Project, WU-Minn Consortium (Principal Investigators: David Van Essen and Kamil Ugurbil; 1U54MH091657) funded by the 16 NIH Institutes and Centers that support the NIH Blueprint for Neuroscience Research; and by the McDonnell Center for Systems Neuroscience at Washington University.



\section*{Appendix}

Abbreviations:  ctx-rh: cortex right-hemisphere  ctx-lh: cortex left-hemisphere

{\small
\begin{longtable}{ | l | l | }
	\hline
	Area name  & No. Of nodes \\ \hline
	\  & \  \\ \hline
	ctx-lh-superiorfrontal & 45 \\ \hline
	ctx-rh-superiorfrontal & 42 \\ \hline
	ctx-rh-precentral & 36 \\ \hline
	ctx-lh-precentral & 35 \\ \hline
	ctx-lh-postcentral & 31 \\ \hline
	ctx-rh-postcentral & 30 \\ \hline
	ctx-lh-superiorparietal & 29 \\ \hline
	ctx-rh-superiorparietal & 29 \\ \hline
	ctx-rh-rostralmiddlefrontal & 27 \\ \hline
	ctx-lh-superiortemporal & 26 \\ \hline
	ctx-lh-rostralmiddlefrontal & 26 \\ \hline
	ctx-rh-inferiorparietal & 26 \\ \hline
	ctx-rh-superiortemporal & 25 \\ \hline
	ctx-rh-lateraloccipital & 23 \\ \hline
	ctx-rh-precuneus & 23 \\ \hline
	ctx-lh-lateraloccipital & 23 \\ \hline
	ctx-lh-precuneus & 22 \\ \hline
	ctx-lh-inferiorparietal & 22 \\ \hline
	ctx-lh-supramarginal & 21 \\ \hline
	ctx-rh-supramarginal & 20 \\ \hline
	ctx-rh-middletemporal & 19 \\ \hline
	ctx-lh-fusiform & 18 \\ \hline
	ctx-rh-lateralorbitofrontal & 17 \\ \hline
	ctx-rh-fusiform & 17 \\ \hline
	ctx-rh-lingual & 17 \\ \hline
	ctx-lh-insula & 17 \\ \hline
	ctx-lh-lingual & 17 \\ \hline
	ctx-lh-inferiortemporal & 16 \\ \hline
	ctx-rh-insula & 16 \\ \hline
	ctx-rh-inferiortemporal & 16 \\ \hline
	ctx-lh-middletemporal & 16 \\ \hline
	ctx-lh-lateralorbitofrontal & 16 \\ \hline
	ctx-lh-caudalmiddlefrontal & 13 \\ \hline
	ctx-rh-paracentral & 12 \\ \hline
	ctx-lh-paracentral & 11 \\ \hline
	ctx-rh-caudalmiddlefrontal & 11 \\ \hline
	ctx-rh-medialorbitofrontal & 11 \\ \hline
	ctx-lh-medialorbitofrontal & 10 \\ \hline
	ctx-lh-parsopercularis & 10 \\ \hline
	ctx-lh-posteriorcingulate & 9 \\ \hline
	ctx-rh-posteriorcingulate & 9 \\ \hline
	ctx-rh-parsopercularis & 9 \\ \hline
	ctx-rh-parstriangularis & 8 \\ \hline
	ctx-rh-cuneus & 8 \\ \hline
	ctx-rh-pericalcarine & 8 \\ \hline
	ctx-lh-cuneus & 7 \\ \hline
	ctx-lh-pericalcarine & 7 \\ \hline
	ctx-lh-isthmuscingulate & 7 \\ \hline
	ctx-lh-parstriangularis & 7 \\ \hline
	ctx-rh-parahippocampal & 6 \\ \hline
	ctx-lh-bankssts & 6 \\ \hline
	ctx-rh-caudalanteriorcingulate & 6 \\ \hline
	ctx-rh-isthmuscingulate & 6 \\ \hline
	ctx-lh-parahippocampal & 6 \\ \hline
	ctx-rh-bankssts & 6 \\ \hline
	ctx-lh-rostralanteriorcingulate & 5 \\ \hline
	ctx-lh-caudalanteriorcingulate & 5 \\ \hline
	ctx-rh-parsorbitalis & 4 \\ \hline
	ctx-lh-transversetemporal & 4 \\ \hline
	ctx-lh-parsorbitalis & 4 \\ \hline
	ctx-rh-rostralanteriorcingulate & 4 \\ \hline
	ctx-lh-entorhinal & 3 \\ \hline
	ctx-lh-temporalpole & 3 \\ \hline
	ctx-rh-temporalpole & 3 \\ \hline
	ctx-rh-transversetemporal & 3 \\ \hline
	ctx-lh-frontalpole & 2 \\ \hline
	ctx-rh-entorhinal & 2 \\ \hline
	ctx-rh-frontalpole & 2 \\ \hline
	Left-Thalamus-Proper & 1 \\ \hline
	Left-Amygdala & 1 \\ \hline
	Right-Hippocampus & 1 \\ \hline
	Right-Amygdala & 1 \\ \hline
	Right-Putamen & 1 \\ \hline
	Right-Accumbens-area & 1 \\ \hline
	Left-Hippocampus & 1 \\ \hline
	Left-Pallidum & 1 \\ \hline
	Right-Pallidum & 1 \\ \hline
	Right-Thalamus-Proper & 1 \\ \hline
	Left-Putamen & 1 \\ \hline
	Right-Caudate & 1 \\ \hline
	Left-Caudate & 1 \\ \hline
	Left-Accumbens-area & 1 \\ \hline
	Brain-Stem & 1 \\ \hline
	\  & \  \\ \hline
	Sum of nodes & 1015  \\ \hline
	 
	\caption*{Table S1: The number of nodes in each ROI.}
\end{longtable}

}

\renewcommand{\thefigure}{S\arabic{figure}}

\setcounter{figure}{0}

\begin{figure}[h!]
	\centering
	\includegraphics[width=150mm]{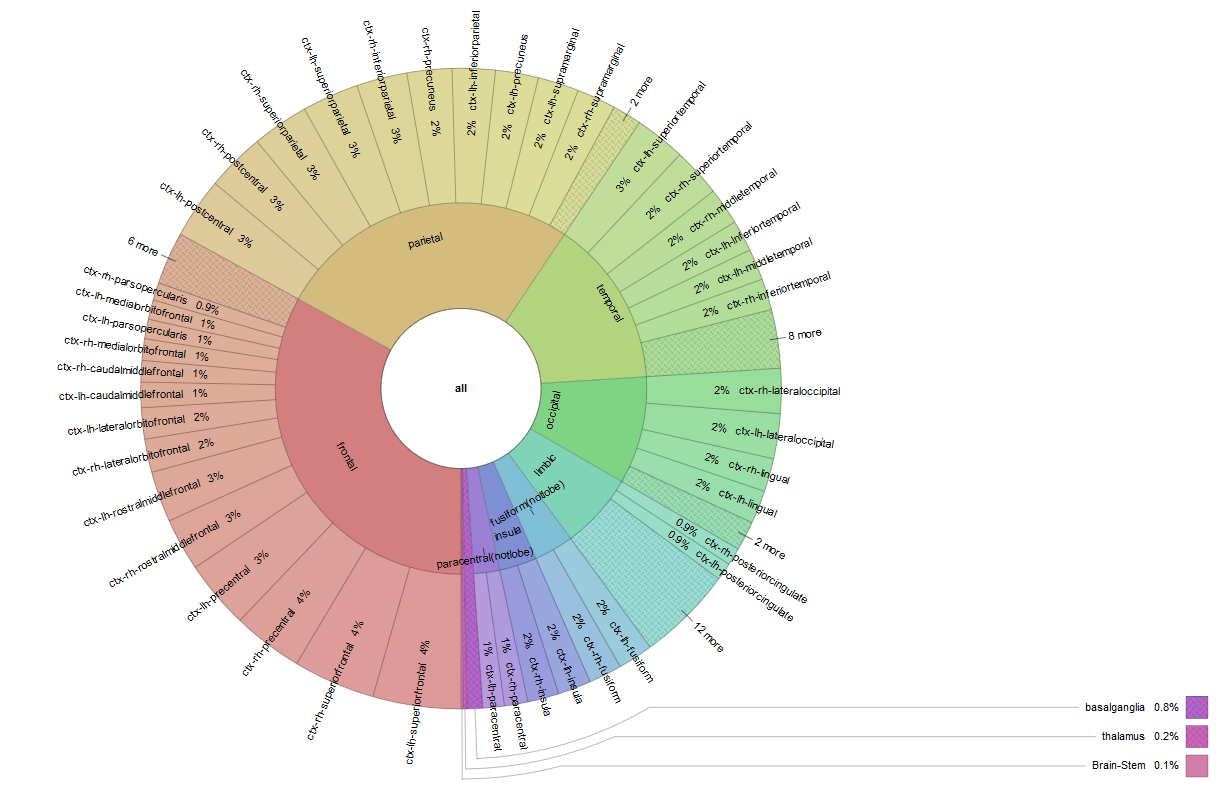}
	\caption{The number of nodes in ROIs and lobes.  The interactive figure can be viewed at \url{http://uratim.com/diversity/Figure_S1-Krona.html}  }
\end{figure}

{\small

\begin{longtable}{ | l | l | }
	\hline
	'all' & 70652 \\ \hline
	'ctx-lh-superiorfrontal' & 910 \\ \hline
	'ctx-rh-superiorfrontal' & 774 \\ \hline
	'ctx-rh-precentral' & 500 \\ \hline
	'ctx-lh-precentral' & 448 \\ \hline
	'ctx-rh-rostralmiddlefrontal' & 352 \\ \hline
	'ctx-rh-inferiorparietal' & 340 \\ \hline
	'ctx-lh-rostralmiddlefrontal' & 331 \\ \hline
	'ctx-lh-superiorparietal' & 317 \\ \hline
	'ctx-rh-superiorparietal' & 314 \\ \hline
	'ctx-lh-postcentral' & 305 \\ \hline
	'ctx-rh-postcentral' & 273 \\ \hline
	'ctx-rh-lateraloccipital' & 263 \\ \hline
	'ctx-lh-lateraloccipital' & 254 \\ \hline
	'ctx-lh-superiortemporal' & 250 \\ \hline
	'ctx-lh-inferiorparietal' & 242 \\ \hline
	'ctx-rh-superiortemporal' & 228 \\ \hline
	'ctx-rh-precuneus' & 227 \\ \hline
	'ctx-lh-precuneus' & 222 \\ \hline
	'ctx-lh-supramarginal' & 209 \\ \hline
	'ctx-rh-supramarginal' & 206 \\ \hline
	'ctx-rh-middletemporal' & 176 \\ \hline
	'ctx-lh-fusiform' & 157 \\ \hline
	'ctx-rh-lateralorbitofrontal' & 144 \\ \hline
	'ctx-lh-inferiortemporal' & 135 \\ \hline
	'ctx-rh-insula' & 131 \\ \hline
	'ctx-lh-lingual' & 131 \\ \hline
	'ctx-rh-fusiform' & 130 \\ \hline
	'ctx-rh-inferiortemporal' & 130 \\ \hline
	'ctx-lh-lateralorbitofrontal' & 127 \\ \hline
	'ctx-lh-insula' & 125 \\ \hline
	'ctx-lh-middletemporal' & 119 \\ \hline
	'ctx-rh-lingual' & 114 \\ \hline
	'ctx-lh-caudalmiddlefrontal' & 91 \\ \hline
	'ctx-rh-paracentral' & 76 \\ \hline
	'ctx-rh-caudalmiddlefrontal' & 65 \\ \hline
	'ctx-lh-paracentral' & 64 \\ \hline
	'ctx-rh-medialorbitofrontal' & 59 \\ \hline
	'ctx-lh-parsopercularis' & 55 \\ \hline
	'ctx-lh-medialorbitofrontal' & 54 \\ \hline
	'ctx-lh-posteriorcingulate' & 45 \\ \hline
	'ctx-rh-parsopercularis' & 45 \\ \hline
	'ctx-rh-posteriorcingulate' & 43 \\ \hline
	'ctx-rh-parstriangularis' & 36 \\ \hline
	'ctx-rh-cuneus' & 35 \\ \hline
	'ctx-rh-pericalcarine' & 35 \\ \hline
	'ctx-lh-cuneus' & 28 \\ \hline
	'ctx-lh-pericalcarine' & 28 \\ \hline
	'ctx-lh-isthmuscingulate' & 28 \\ \hline
	'ctx-lh-parstriangularis' & 28 \\ \hline
	'ctx-lh-bankssts' & 21 \\ \hline
	'ctx-rh-caudalanteriorcingulate' & 21 \\ \hline
	'ctx-lh-parahippocampal' & 21 \\ \hline
	'ctx-rh-parahippocampal' & 20 \\ \hline
	'ctx-rh-isthmuscingulate' & 20 \\ \hline
	'ctx-rh-bankssts' & 20 \\ \hline
	'ctx-lh-rostralanteriorcingulate' & 15 \\ \hline
	'ctx-lh-caudalanteriorcingulate' & 15 \\ \hline
	'ctx-rh-parsorbitalis' & 10 \\ \hline
	'ctx-lh-parsorbitalis' & 10 \\ \hline
	'ctx-rh-rostralanteriorcingulate' & 10 \\ \hline
	'ctx-lh-transversetemporal' & 8 \\ \hline
	'ctx-lh-entorhinal' & 6 \\ \hline
	'ctx-rh-transversetemporal' & 5 \\ \hline
	'ctx-lh-temporalpole' & 4 \\ \hline
	'ctx-rh-entorhinal' & 3 \\ \hline
	'ctx-rh-temporalpole' & 3 \\ \hline
	'Left-Thalamus-Proper' & 1 \\ \hline
	'Left-Amygdala' & 1 \\ \hline
	'ctx-lh-frontalpole' & 1 \\ \hline
	'Right-Hippocampus' & 1 \\ \hline
	'Right-Amygdala' & 1 \\ \hline
	'ctx-rh-frontalpole' & 1 \\ \hline
	'Right-Putamen' & 1 \\ \hline
	'Right-Accumbens-area' & 1 \\ \hline
	'Left-Hippocampus' & 1 \\ \hline
	'Left-Pallidum' & 1 \\ \hline
	'Right-Pallidum' & 1 \\ \hline
	'Right-Thalamus-Proper' & 1 \\ \hline
	'Left-Putamen' & 1 \\ \hline
	'Right-Caudate' & 1 \\ \hline
	'Left-Caudate' & 1 \\ \hline
	'Left-Accumbens-area' & 1 \\ \hline
	'Brainstem' & 1 \\ \hline
\caption*{Table S2: The number of edges in each ROI.}
\end{longtable}

}


\end{document}